# *Ab initio* vacancy formation energies and kinetics at metal surfaces under high electric field

Shyam Katnagallu[1,*], Christoph Freysoldt[1,*], Baptiste Gault[1,2] and Jörg Neugebauer[1]

[1] Max Planck Institut für Eisenforschung GmbH, Max-Planck- Str. 1, 40237, Düsseldorf, Germany.

[2] Royal School of Mines, Imperial College, Prince Consort Road, SW7 2BP, London, UK.

* Corresponding authors: s.katnagallu@mpie.de; c.freysoldt@mpie.de

**Abstract:**

Recording field ion microscope images under field evaporating conditions and subsequently reconstructing the underlying atomic configuration, called three-dimensional field ion microscopy (3D-FIM) is one of the few techniques capable of resolving crystalline defects at an atomic scale. However, the quantification of the observed vacancies and their origins are still a matter of debate. It was suggested that high electric fields (1–5 V/Å) used in 3D-FIM could introduce artefact vacancies. To investigate such effects, we used density functional theory (DFT) simulations. Stepped Ni and Pt surfaces with kinks were modelled in the repeated slab approach with a (971) surface orientation. An electrostatic field of up to 4 V/Å was introduced on one side of the slab using the generalized dipole correction. Contrary to what was proposed, we show that the formation of vacancies on the electrified metal surface is more difficult compared to a field-free case. We also find that the electric field can introduce kinetic barriers to a potential "vacancy-annihilation" mechanism. We rationalize these findings by comparing to insights from field evaporation models.

**Introduction**

Crystal defects such as vacancies, impurities, dislocations, stacking faults, and grain boundaries significantly modify the mechanical, functional, or chemical properties of crystalline materials [1]. Mechanical or thermal processing steps create, alter, or annihilate such defects and hence allow for fine tuning a material's performance in principle. Yet, the precise relationships between processing-microstructure-properties remain unpredictable in most cases due to a lack of detailed understanding the mutual interplay of different defects. Among such defects, the analysis of individual isolated vacancies remains broadly inaccessible despite their importance in material phenomena such as creep, dislocation motion and pinning, and diffusional transport. Populations of vacancies and their interactions with neighbouring solutes can be indirectly characterized by e.g. differential dilatometry and positron life-time annihilation spectroscopy [2]. These techniques require a $10^{15}$-$10^{19}$ cm$^{-3}$ concentration of vacancies [3] in the material, thus limiting the estimation of vacancy concentrations encountered at high temperatures after quenching. The atomistic details of the structural and chemical interactions of the vacancies in materials can only be indirectly inferred. Many efforts have been pursued to achieve true atomic resolution, for instance through field ion microscopy (FIM) [4] and through advances in atomic electron tomography [5,6].

FIM was the first technique to image individual atoms on a metal surface [7] and it represents an alternative to electron-based techniques. FIM has helped to understand the structure of crystalline defects such as grain boundaries, stacking faults, dislocations and even vacancies [8–12]. FIM relies on ionization of an inert imaging gas near the surface of specimen which is kept at electrostatic fields in the range of 1–5·$10^{10}$ V/m. Such intense electric fields are achieved experimentally by shaping the specimen as a sharp needle with an end radius smaller than 100 nm. Needle specimens from crystalline materials often exhibit stepped terraces where major sets of crystallographic planes intersect the curved surface [13]. In FIM, imaging of certain terraces is atomically resolved, due to atomic scale corrugation of the electric field [14,15], and thus allowing to image vacancies. Recent developments have led to 3D-FIM, which involves recording the field ion images of a continuously field-evaporating specimen. Image processing algorithms deployed on the series of micrographs allow a full 3D reconstruction [16], including of individual atomic positions [17–19], thereby enabling quantification of the distribution and interaction of vacancies with solute elements [18,20]. Although 3D-FIM has the potential to estimate even small vacancy concentrations along with their spatial distribution, this capability has been questioned previously.

Early studies [12,21,22] showed that the vacancy concentration in some metals was much higher than the predicted equilibrium vacancy concentration. This led to the conclusion that most of the vacancies observed in FIM were artefacts that can originate from: (i) field-induced etching in presence of certain gases in the chamber; (ii) preferential field evaporation of solute atoms; (iii) reduction of vacancy formation energy at the surface due to the high electric field; and (iv) bombardment of the metal surface by polarized imaging atoms. Among these proposed mechanisms, the first two can be avoided by precisely controlling the experimental environment, i.e., by using high purity imaging gas and ensuring the purity of the specimen. The fourth mechanism has only been suggested to occur in metals like gold [22], but not in refractory metals.

Here, in order to investigate the influence of the electric field on the vacancy formation energy, we used density functional theory (DFT) calculations. Kohn Sham-DFT [23] forms the basis of modern day ab initio electronic structure calculations, and is widely applied in various fields of science [24]. To efficiently apply electric field in periodic boundary conditions we use the generalized dipole correction method [25,26]. Stepped Ni and Pt surfaces with kinks were modelled with a (971) surface orientation. An electrostatic field of up to 4 V/Å was introduced on one side of the slab using the generalized dipole correction. We find that the vacancy formation energy *increases* with application of electric field in both Ni and Pt. We also find the creation and suppression of energy barriers of atomic movements, under application of field. We show that such calculations can improve the confidence in extracting the vacancy concentrations and atomic environments from 3D-FIM.

**Methods:**

On the surface, steps and kinks are the relevant sites for field ion imaging and field evaporation due to their lower coordination. To represent such features using periodic slab models that are computable within DFT (100-1000 atoms), carefully chosen high-index surface orientations are employed. To setup the simulations we designed a tool to generate metallic slabs of high Miller index with selected terrace, length and direction of steps and kinks, as shown in Figure 1 (a) , based on the work by Van Hove et al. [27]. This tool and all DFT calculations we discuss below are set up and analysed in pyiron [28], which is an integrated development environment for atomistic simulations. We created a face-centered cubic (fcc) Ni (971) surface (66 atoms), which has a (100) type terrace with 5 atoms long steps and 1 atom long kink. The slab is shown in Figure 1(b).  In order to apply an electric field on one side of the slab we use the generalized dipole correction implemented in S/PHI/nX [25,29].

For the calculations we use the projector augmented wave (PAW) pseudopotentials taken from the VASP library and PBE [30] exchange correlation functional. We ensured a careful convergence with respect to the **k**-points and energy cut-off for plane waves, to yield an accuracy of $10^{-3}$ eV in the relevant energies. A **k**-point sampling of 6x6x1 and a plane-wave energy cut-off of 500 eV was used.

Application of electric field usually induces a tensile Maxwell stress on the slab in the direction of the field. This requires then an additional convergence criterion, which was set with respect to the total thickness of the slab (9 Å) and height of the fixed layers (4 Å) in the slab to yield converged interlayer relaxations to within 0.1%. Similarly, for Pt (971) a slab (80 atoms) with height (13 Å), an energy cut-off of 400 eV with a **k**-point sampling of 6x6x1 was used. Additional care must be taken for electronic occupancy smearing methods and widths used, as the charge distributions under these high electric fields proved to be sensitive to them. We used Fermi-Dirac smearing with first order entropy corrections with a width of 0.1 eV.

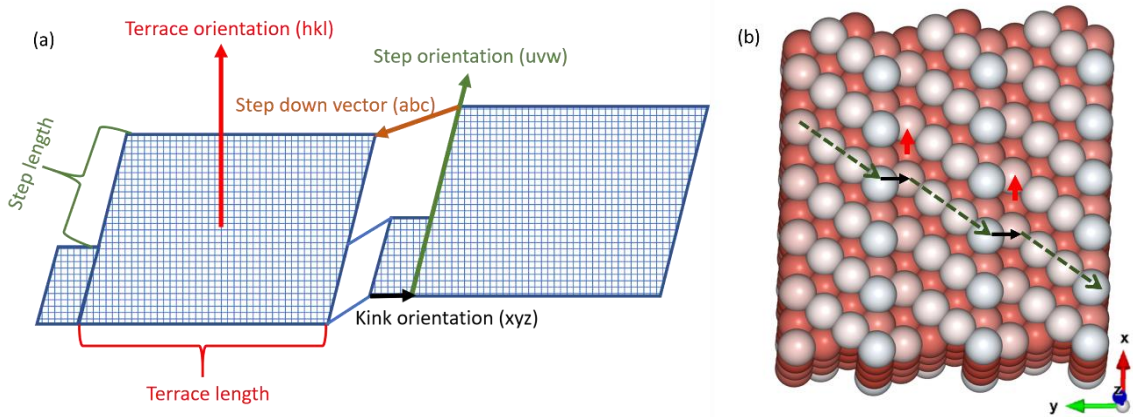

*Figure 1: a) Schematic drawing of the high miller index surface, decomposed into relevant terrace, step and kink orientations and their lengths. b) Ni (971) surface composed of (100) type terraces indicated by red arrows, with steps and kinks indicated with dashed green and black arrows respectively. The atoms are coloured according to the number of first nearest neighbours.*

To evidence the influence of the electric field, the vacancy formation energies ($E_f$) in Ni and Pt were calculated as a function of their lattice position within the slab. These values are obtained using the following equation:

$$E_f = E_{\text{vac}} - E_{perf} + E_{\text{bulk}} \qquad 1$$

where $E_{\text{vac}}$ is the total energy of the cell containing a vacancy, $E_{\text{perf}}$ is the energy of the perfect slab without any defects under same conditions, and $E_{\text{bulk}}$ is the energy per atom in a bulk cell. The vacancy formation energies are also calculated for the same slabs without any electric field.

**Results and Discussion:**

The vacancy formation energies are plotted as a function of the vacancy position within the slab, where the right end (10 Å and 14 Å) of the slab corresponds to the surface and the left to bulk, with and without an applied electric field of 4 V/Å for Ni and Pt, in Figure 2 (a) and (b) respectively. For both metals, the vacancy formation energies are higher when the electric field is applied, i.e., the electric field does not facilitate the formation of vacancies. This can likely be attributed to the induced Maxwell-stress combined with the resulting material's mechanical response, which makes it difficult for vacancies to form compared to the field-free case. For Pt, the large site-to-site variations both with and without electric field are related to the charge density redistributions occurring, even in absence of geometrical relaxation, rather than due to unique structural relaxations for different vacancy positions (see Supplementary Figure S1).

The vacancy formation energy in the bulk-like regions of the slab is compared with experimental and other DFT calculations in literature. In this study, for Ni the vacancy formation energy near the bulk-like region of the slab without electric field is 1.48 eV. Using a PBE functional, Nandi et.al. [31], have reported and 1.42eV for Ni bulk and experimentally reported value through positron annihilation [32] is 1.73 ± 0.07 eV. For Pt, the vacancy formation energy in the bulk like region our slab without electric field is 0.80 eV. Mattsson et.al., [33] reported a value of 0.68 eV as vacancy formation energy using PBE for Pt bulk. Whereas, the experimental reported values range from 1.24-1.45 eV [34] for Pt vacancy formation energy. The systematic discrepancy between the experimental and the *ab initio* calculated values could arise from the nonlinear temperature dependence of Gibbs formation energy of vacancies, which are not corrected in experimental data when extrapolating to 0K [35], or from deficiencies in the functionals to reproduce accurate surface energies [36].

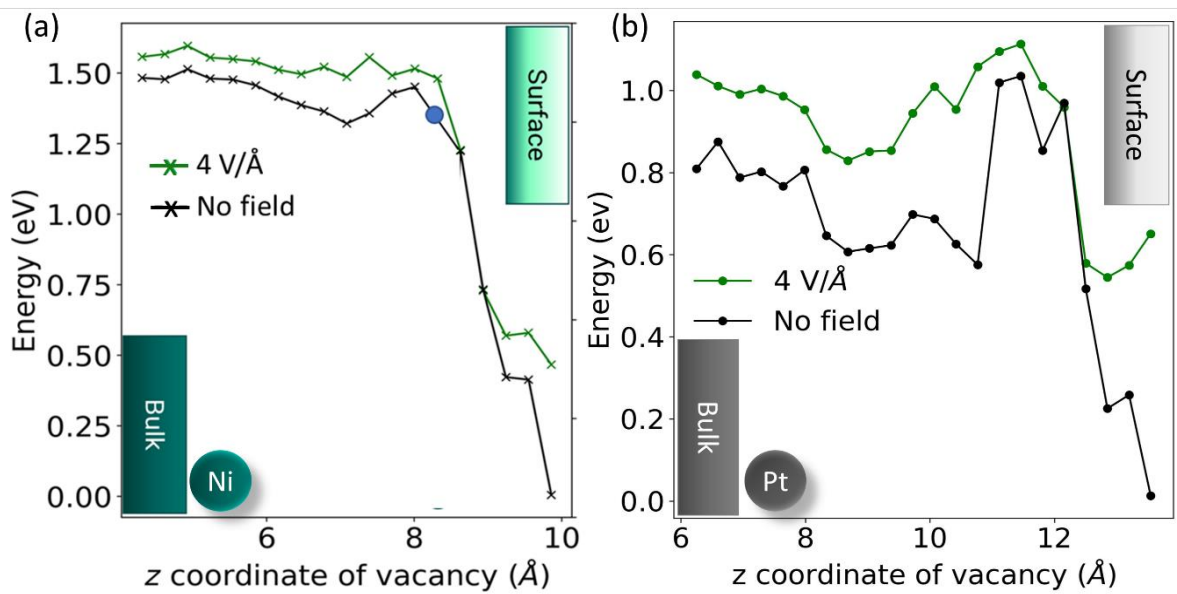

*Figure 2: Vacancy formation energy vs the position of the vacancy in the slabs with (green) and without (black) an electric field in (a) Ni and (b) Pt. The point of where vacancy spontaneously moves to kink position is marked with a blue solid circle in (a).*

The Ni vacancy formation energy profiles in Figure 2(a) exhibit a constant offset of ~0.085 eV ± 0.01 eV in the bulk-like region of the slab. The higher vacancy formation energy under field can be attributed to the tensile strain from the Maxwell stress $\sigma_{zz} = \frac{1}{2}\epsilon_0 E_z^2 \approx 7$ GPa induced by the electric field $E_z$. To verify this hypothesis, we approximated the total strain in the slab as a uniaxial tensile strain extracted from the displacements of atoms in the slab along the z-axis (applied field direction). The average displacement along z as a function of the z coordinate of atoms is shown in Figure S2 (a). This average strain (3.49%) is then applied to a 66-atom bulk Ni supercell oriented along the same high Miller index surface (971) direction used for creating the slabs. The difference in bulk vacancy formation energy under strain and strain free case is 0.11 eV, which is close to the 0.085eV offset in our calculations. The volumetric effect of the vacancy is exemplified in Fig S2 (b). It shows that vacancy formation energy increases, as a function of uniaxial tensile strain applied to the bulk Ni. This indicates that the defect's excess volume tensor's interaction with the stress can in fact increase the vacancy formation energy. This effect would be similar for vacancies across all metals, and thus in general application of electric field would make it difficult to form vacancies. Similar stress effects should be expected for other defects as well and scale with the defect's excess volume (also called 'λ') tensor [37].

We then focus on a configuration for Ni in which a vacancy is positioned in a layer below the kink site, indicated by the blue circle in Fig 2 (a) and depicted in the inset in Figure 3(a), with the considered Ni atom at the kink site in orange, and vacancy position as a white circle. The trajectory of the atom during its relaxation in the field free case, Figure 3(a), has no discernible barrier. This vacancy-kink atom swap annihilates the vacancy, but retains the same high index configuration as when the vacancy is on the surface when no electric field is applied. However, upon application of a 4 V/Å electric field, as shown in Figure 3 (b), an energy barrier of approx. 0.15 eV appears as the kink atom is moved towards the vacancy position following the same trajectory. We also plotted the mean displacements of the kink atom and the atom sitting next to the kink atom on the step (marked in purple in the inset). This shows that only the kink atom is involved in the vacancy swapping. The energy barrier prevents the spontaneous vacancy annihilation from readily occurring. Additionally, the barrier for field evaporation would be very small and the kink atom is more likely to evaporate under such conditions, rather than in participating in vacancy annihilation. It originates from the lower electron density of the kink atom, a consequence of its lower coordination, making it subject to a higher outward pulling force from the electric field. The impact of the electric field on the barrier height in the present case is analogous to the case of field evaporation [26], but with opposite sign. In field evaporation, a positively charged surface atom moves towards the negatively charged counter electrode, and barriers generally *decrease* with increasing electric field. In subsurface vacancy annihilation, the atomic movement is in the opposite direction, and hence barriers *increase*.

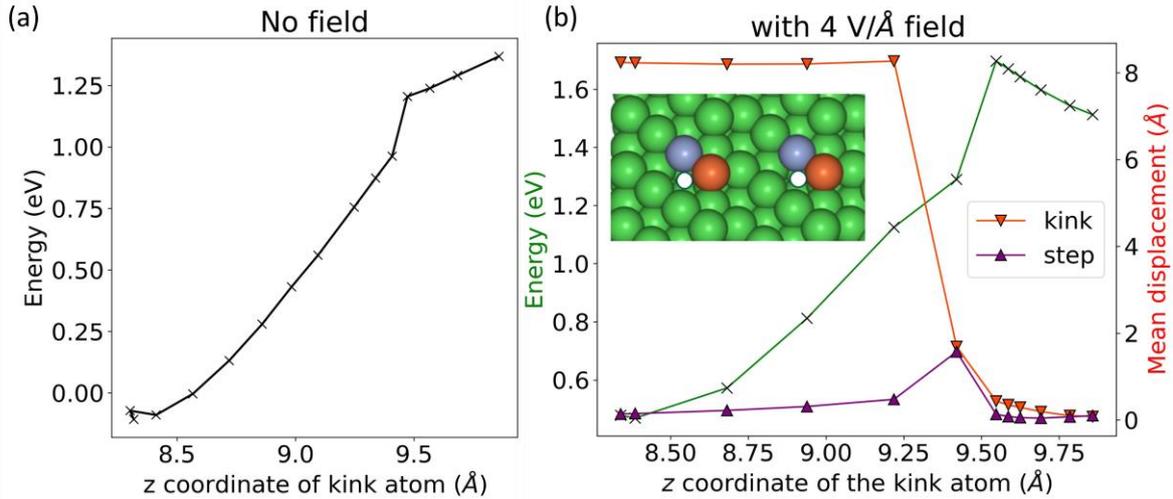

*Figure 3: (a) Potential energy landscape for a kink atom along the path to a vacancy position. Note that the kink atom can move into the subsurface vacancy without having to overcome a barrier. (b) Potential energy landscape for the same path in the presence of an electric field. The inset shows the slab's atomic configuration repeated twice, with the kink atom coloured orange, the step atom coloured purple and the vacancy position indicated by a white solid circle. The graph also shows the mean displacement of the kink atom and of the neighbouring step atom.*

Next, to investigate the effect of a solute atom on vacancy annihilation, we followed the same barrier calculations, now with a Ta atom at the kink site of the Ni slab. We chose Ta, as in our recent 3D-FIM study [18] Ta was shown to be bound with a vacancy in a creep deformed Ni-2 at% Ta alloy. DFT calculations with this configuration showed that the Ta atom also drops down to the vacancy position without the electric field. Interestingly, when a 4 V/Å field is applied in the calculations, Ta stays at the kink site, but the Ni step atom next to the Ta drops down to the vacancy position in contrast to our observation in pure Ni under field. This can be attributed to the presence of an energy barrier (0.3 eV) for the step Ni atom next to the Ta atom, when no electric field is applied. This barrier for the step atom is calculated by fixing the Ta atom and is shown in Figure S3.

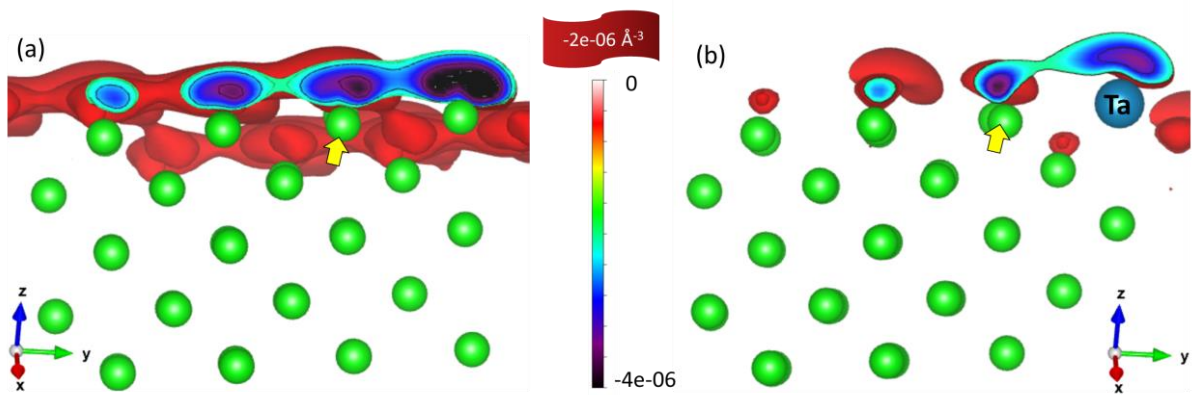

*Figure 4: -2e-06 Å$^{-3}$ iso surfaces corresponding to an electron density difference due to application of a field in a tilted (a) Ni (971) slab and (b) Ni-Ta (971) slab. Cross sections through the plane containing the step and kink atom in Ni (971) and Ni-Ta (971) slab are also shown in the iso surfaces. The position of the step atom is indicated by a yellow arrow, which shows that the step atom gains a smaller net positive charge, when Ta is present next to it.*

This contrasting behaviour in the presence of Ta, can be explained by the electronic charge density difference for a Ni slab with and without field, and with addition of a Ta solute. Figure 4 (a) and (b) show the iso-surfaces corresponding to $-2.0*10^{-6}$ Å$^{-3}$ electron density difference, which corresponds to a net positive charge, for Ni and Ni-Ta slab respectively. It can be seen that the step atom, indicated by the purple arrow, develops a higher and wider net positive charge in the pure Ni compared to when it is neighbouring the Ta atom. This is further illustrated through the sections in electron density difference taken along planes containing the position of step and kink atoms, shown in Figure 4 (a) and (b) for the Ni and Ni-Ta slabs, respectively. Only negative electron density differences (i.e., positive charges) are plotted for simplicity. The difference in the net positive charge gained by the step atom, based on its chemical environment, can be attributed to the atom's electronegativities: Ni has an electronegativity of 1.9, i.e. higher than 1.5 for Ta [38]. Thus, when Ta is present, Ni tends to have a higher electron density (see Figure S4) and can only gain a smaller net positive charge when the electric field is applied. This in turn leads to a smaller force acting on the atom and so it can swap its position with the vacancy, unlike when a Ni atom is present at the kink site.

In summary, applying an intense electric field generally increases the energy of vacancy formation in Ni and Pt, making the creation of artefact vacancies in FIM unlikely. The Maxwell stress generated by the electric field slightly increases the vacancy formation energies also in the bulk. In configurations where a vacancy can spontaneously move to the surface, the application of the electric field generates an energy barrier that can inhibit the exchange of the position with a surface atom. This mechanism for vacancy annihilation on the surface remains possible when a solute, here the more electropositive Ta, is in a kink site because of the decrease in charge density on the Ni atom. However, provided these extracted vacancy concentrations from FIM are statistically significant, as facilitated by developments of the FIM data extraction routines [19,39], these annihilations are unlikely to cause a substantial drop. This analysis can further be extended to other metal systems, based on the vacancy excess volume tensor and its interaction with the stress due to electric field. These simulations can improve the confidence of the extracted vacancy concentrations and structures through 3D-FIM.

**Acknowledgements:**

SK and CF thank the BIGmax consortium, the Max Planck Society's Research Network on Big-Data-Driven Materials Science for funding this work. SK and BG are grateful for funding from the DFG through the Leibniz Prize 2020.

**Supplementary:**

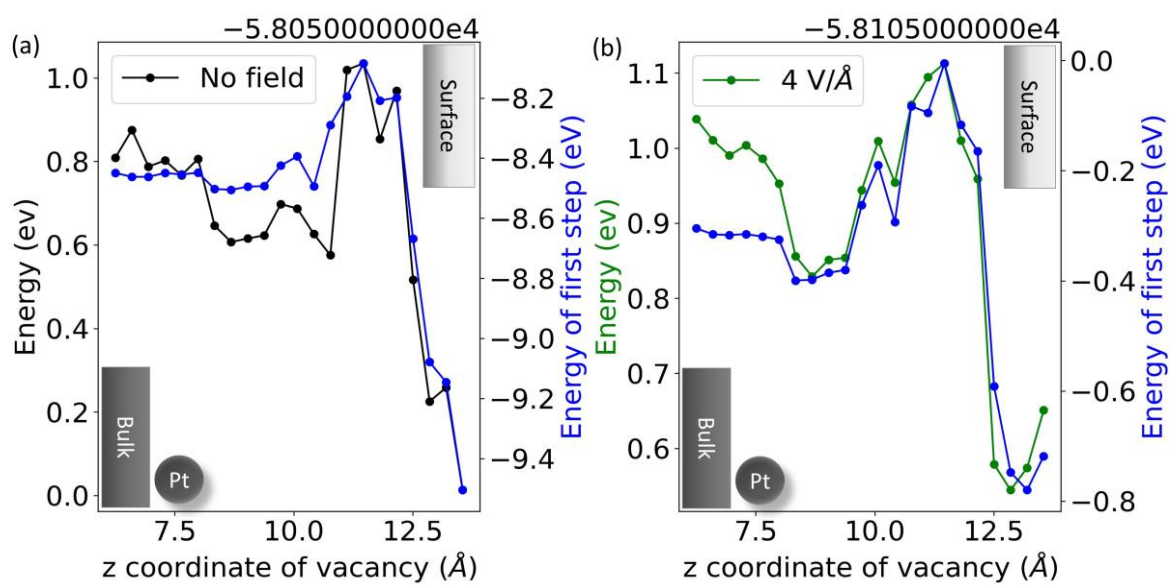

*Figure S 1: Vacancy formation energy as a function of the vacancy position for (a) without electric field and (b) with 4 V/Å electric field.*

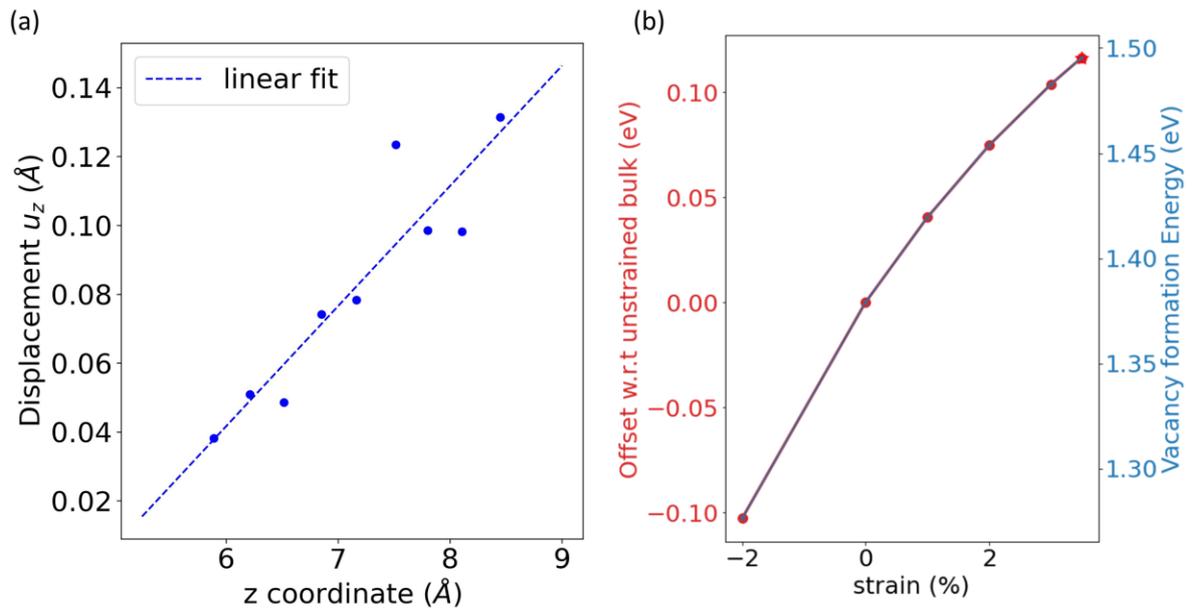

Figure S 2: (a) The average displacement of atoms along the z axis (the applied field direction) of the slab as a function of their z coordinate. A linear fit assuming the tensile strain caused by the Maxwell stress from 4 V/Å field is also plotted. (b) vacancy formation energies in bulk Ni cells subjected to different strains. The offset with respect to unstrained bulk is plotted on the left axis and absolute values on the right.

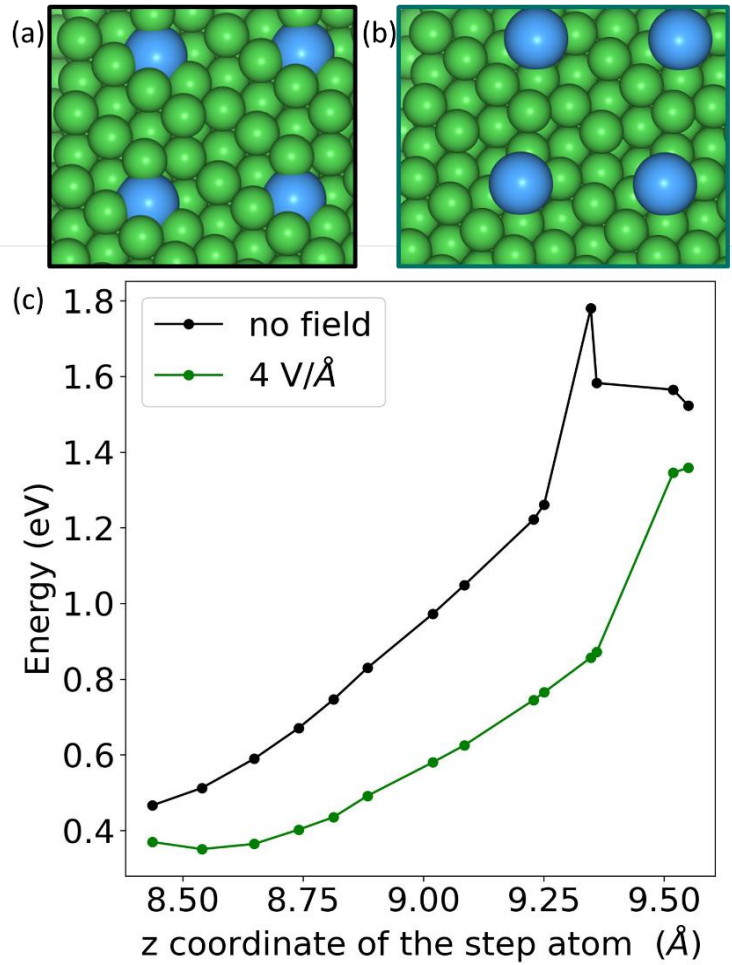

Figure S 3: (a) Final configuration of Ta swapping the position with a vacancy when no field is applied. (b) Final configuration of a Ni step atom swapping its position with a vacancy under applied electric field. (c) Barrier faced by the step atom with and without field.

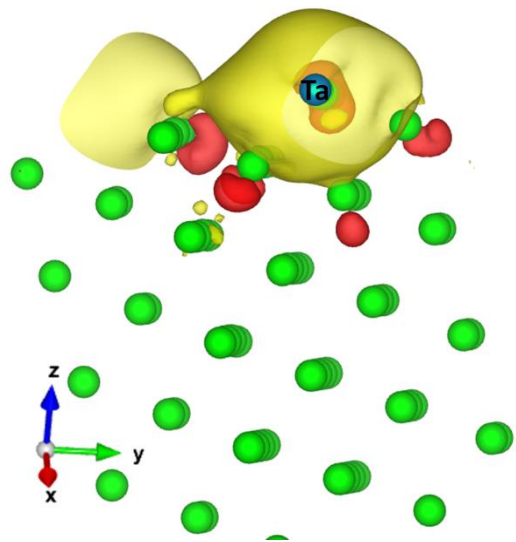

Figure S 4: Electron density difference between Ta at the kink position and Ni at kink site. The yellow iso surface corresponds to a positve electron density difference of 9.5e-07 Å$^{-3}$ and red iso surface to the negative electron density difference of sasme value. The presence of Ta creates an additional electron density at the nearest Ni neighbours.